\documentclass[12pt, a4paper, fleqn]{article}
\usepackage{amsmath}
\usepackage{amssymb}
\usepackage{graphicx}
\usepackage{caption}
\usepackage{adjustbox}
\usepackage{fancyhdr}
\usepackage{lmodern}
\usepackage{placeins}
\usepackage{appendix}
\usepackage{bbm}
\usepackage{color}
\usepackage[normalem]{ulem}

\usepackage[left=2cm,right=2cm,top=3cm,bottom=2.5cm]{geometry}
\usepackage[utf8]{inputenc}

\title{\vspace{-1cm} Finite-horizon Linear Quadratic Control for Networked Control Systems with non-distributed plants}

\renewcommand\footnotemark{}

\author{Marijan Palmisano $(1)$, Martin Steinberger $(1)$, Martin Horn $(1,2)$% <-this % stops a space
	\thanks{{\small $(1)$ Marijan Palmisano, Martin Steinberger and Martin Horn are with the Institute of Automation and Control, Graz University of Technology, Graz, Austria. 
		E-mail: {\tt\footnotesize  marijan.palmisano@tugraz.at}, {\tt\footnotesize martin.steinberger@tugraz.at}, {\tt\footnotesize martin.horn@tugraz.at}.}}
	\thanks{{\small $(2)$ Martin Horn is with the Christian Doppler Laboratory for Model Based
			Control of Complex Test Bed Systems, Institute of Automation and Control, Graz University of Technology, Graz, Austria.\newline
			The financial support by the Christian Doppler Research Association, the
			Austrian Federal Ministry for Digital and Economic Affairs and the National
			Foundation for Research, Technology and Development is gratefully
			acknowledged.}}% <-this % stops a space
} 
\date{}

\begin{document}
	\maketitle
	
	\begin{abstract}
	An optimal control law for networked control systems with a discrete-time linear time-invariant (LTI) system as plant and networks between sensor and controller as well as between controller and actuator is proposed. This controller is designed by solving an optimization problem that is a generalization of the optimization problem used to obtain the Linear Quadratic Regulator (LQR) for deterministic discrete-time LTI systems. The networks are represented by random delays and drop outs of transmitted data packets.
	\end{abstract}

	\section{Introduction}
	Networked control systems (NCS) where the controllers, actuators and sensors are connected via networks come with several advantages but also with new challenges compared to classical control strategies, see \cite{Zhang2017} and \cite{Zhang2020}. Two major challenges are random delays and drop-outs of transmitted data packets which are both accounted for in the proposed control law.
	
	An optimal control law is designed for a closed control loop that consists of a discrete-time linear time-invariant (LTI) system as plant and two networks, one between sensor and controller and one between controller and actuator. While this setup looks similar to the closed control loop for ``TCP like protocols'' in \cite{Schenato2007}, there are several differences:
	\begin{itemize}
		\item Data can not only be randomly dropped out but can also be randomly delayed.
		\item The information about data packets received by the network interface of the actuator is not obtained via ``TCP like protocols'' but is sent to the controller together with the measured state vector. This allows the use of other protocols but requires a non-distributed plant (in the sense that data can be transferred form the network interface of the actuator to the network interface of the sensor within one time step).
		\item The actuating variable is always selected from the data contained in the most recent packet available at the network interface of the actuator instead of applying zero in case of data drop-out.
		\item The entire state vector is measured and noise is not considered.
	\end{itemize}
	
	Plant and networks together can also be represented by an equivalent jump linear system using and extended state vector containing previous values of the input variable. Control laws that require this extended state vector like in \cite{Shousong2003} can not be applied to our control loop since this vector is not generally available at the controller.
	
	Another approach to the design of control laws for similar control loops is for example taken in \cite{Wang2010} and \cite{Li2019} where a linear state controller is parameterized such that stability of the closed control loop can be guaranteed. There are also approaches where plant and networks are first written as jump linear system followed by choosing parameters of a specific class of control laws such that the closed control loop is stable like in \cite{Xiao2000} (mode-independent dynamic feedback controllers) and \cite{Blind2008} (mode-dependent linear state controllers). 
	
	Other approaches utilize buffers in order to handle random delays like for example in \cite{Repele2014}, \cite{Ludwiger2017}, \cite{Ludwiger2018} and \cite{Ludwiger2019}. Such buffers induce additional delay and are not part of the proposed control law.

%%%%%%%%%%%%%%%%%%%%%%%%%%%%%%%%%%%%%%%%%%%%%%%%%%%%%%%%%%%%%%%%%%%%%%%%%%%%%%%%
\section{Problem formulation}
\subsection{Control loop}\label{sec:loop}
\begin{figure}[h]
	\centering
	\includegraphics[width=0.7\textwidth]{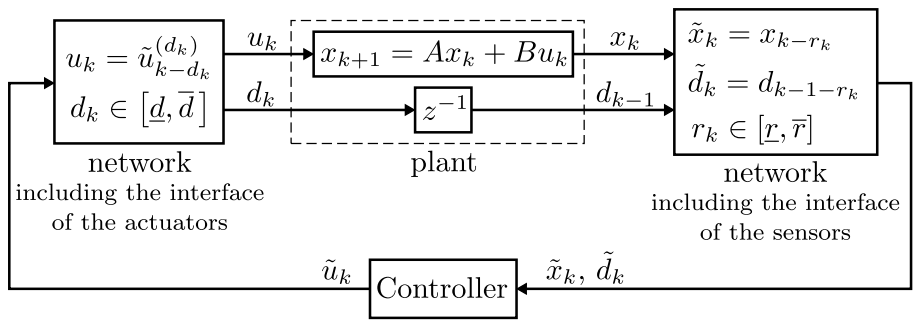}
	\caption{Closed control loop at time step $k$}
	\label{fig:1}
\end{figure}

The closed control loop consisting of a discrete-time LTI system
\begin{align}
x_{k+1} &= Ax_k + Bu_k\label{eq_I_1}
\end{align}
with the state vector $x_k\in\mathbb{R}^n$ and the actuating variable $u_k\in\mathbb{R}^m$ as plant, the controller and two networks is shown in Figure~\ref{fig:1}.

The networks are synchronized so the delay of a received packet is known. Random delays and drop-outs are taken into account via the discrete-time stochastic processes $\mathbf{\mathcal{R}} = (\mathcal{R}_{k_0},\mathcal{R}_{k_0 + 1},...)$ and $\mathbf{\mathcal{D}} = (\mathcal{D}_{k_0},\mathcal{D}_{k_0 + 1},...)$ with the corresponding realizations $\mathbf{r} = (r_{k_0},r_{k_0 + 1},...)$ and $\mathbf{d} = (d_{k_0},d_{k_0 + 1},...)$. The realization $r_k$ of the random variable $\mathcal{R}_k$ is the number of time steps that have passed since the most recent packet available at the controller at time step $k$ has been transmitted, the realization $d_k$ of the random variable $\mathcal{D}_k$ is the number of time steps that have passed since the most recent packet available at the network interface of the actuator at time step $k$ has been transmitted. The most recent packet available at time step $k$ is not necessarily the most recently received packet since packets can overtake each other.

It is assumed that the delays and the number of successive drop-outs are bounded such that $r_k$ and $d_k$ are also bounded by
\begin{alignat}{3}
r_k&\in\left[\underline{r},\ \overline{r}\right],\quad &&0\leq\underline{r}\leq\overline{r},\quad &&r_k,\underline{r},\overline{r}\in \mathbb{N}_0\\
d_k&\in\left[\underline{d},\ \overline{d}\right],\quad &&0\leq\underline{d}\leq\overline{d},\quad &&d_k,\underline{d},\overline{d}\in \mathbb{N}_0.\label{eq_bound_d}
\end{alignat}

The state variable $x_k\in\mathbb{R}^n$ is measured and sent to the controller together with $d_{k-1}$ as part of the same data packet. The controller sends 
\begin{align}
\tilde{u}_k&=\begin{bmatrix}\tilde{u}_k^{(\overline{d})}\\ \vdots\\ \tilde{u}_k^{(\underline{d})}\end{bmatrix}\label{eq:P_3}
\in\mathbb{R}^{\tilde{m}}
\end{align}
where
\begin{align}
\tilde{m} = (\overline{d}-\underline{d}+1)m\label{eq4_1}
\end{align}
to the plant, the actual actuating variable $u_k \in\mathbb{R}^{m}$ is selected at the network interface of the actuator as 
\begin{align}
u_k &= \tilde{u}^{(d_k)}_{k-d_k} \quad\Rightarrow\quad u_{k+p} = \tilde{u}^{(d_{k+p})}_{k+p-d_{k+p}}\label{eq:2_1}
\end{align}
so $\tilde{u}^{(p)}_{k}$ is applied as actuating variable $u_{k+p}$ if $d_{k+p}=p$ since
\begin{align}
u_{k+p}\big|_{d_{k+p}=p} &= \tilde{u}^{(d_{k+p})}_{k+p-d_{k+p}}\bigg|_{d_{k+p}=p} = \tilde{u}^{(p)}_{k}.\label{eq_I_4}
\end{align}
If $d_{k+p}>p$ the data packet containing $\tilde{u}_k$ has not yet been received at time step $k+p$ and if $d_{k+p}<p$ a more recent packet has already been received at time step $k+p$ so the actual actuating variable $u_{k+p}$ is selected from the values contained in that packet according to \eqref{eq:2_1}.

\subsection{Information set}
The information set
\begin{align}
\mathcal{I}_k &= \big\{ (r_k,r_{k-1},...),(\tilde{x}_k,\tilde{x}_{k-1},...),(\tilde{d}_k,\tilde{d}_{k-1},...), (\tilde{u}_{k-1},\tilde{u}_{k-2},...)\big\}\label{eq_I_5}
\end{align}
is available at the controller at time step $k$ where
\begin{align}
\tilde{x}_k &= x_{k-r_k} \label{eq:Ar3}
\end{align}
is the most recent value of $x$ and
\begin{align}
\tilde{d}_k &= d_{k-1-r_k}
\end{align}
is the most recent value of $d$ available at the controller.

\subsection{Transition probabilities}
Both $\mathcal{R}$ and $\mathcal{D}$ are assumed to be stationary Markov processes as described in detail in \cite{Meyn1993}. Therefore if $d_{k-c} = a$ with $c\in\mathbb{N}_0$ is the most recent available value of $\mathbf{d}$, the probability for $d_{k+p} = b$ with $p\in\mathbb{N}_0$ only depends on $d_{k-c}$. This transition probability is written as
\begin{align}
\Phi_{(p+c)}(a,b) &= \mathbb{P}\left(d_{k+p}=b \,|\, d_{k-c}=a\right). \label{eq2_11}
\end{align}
The same applies to $r$ but only the probabilities
\begin{align}
\Psi(a,b) &= \mathbb{P}\left(r_{k+1}=b \,|\, r_k=a\right)\label{eq:P_7}
\end{align}
are used for further calculations.

The number of time steps that have passed since the most recent available packet has been sent can not increase by more than $1$ each step, i.e. $d_{k+p} \leq d_k+p$ and $r_{k+p} \leq r_k + p$. Additionally it is reasonable to assume that for each possible delay there is a non-zero probability of receiving a packet with the respective delay. Therefore
\begin{align}
\Psi(a,b) > 0\ \Leftrightarrow b\leq a+1 \label{eq:Ar13}
\end{align}
with $a,b\in\left[\underline{r},\ \overline{r}\right]$ and
\begin{align}
\Phi_{(p+c)}(a,b) > 0\ \Leftrightarrow b\leq a+p+c \label{eq3_9}
\end{align}
with $a,b\in\left[\underline{d},\ \overline{d}\right]$.

\section{Controller design}
\subsection{Optimization problem}\label{sec:problem}
The optimization problem that is solved in order to obtain an optimal control law $\tilde{u}_k$ for $k = k_0,...,N$ is given by $\mu_{k_0}\left(J\right)$ where
\begin{align}
\mu_{N+1}\left(J\right) &= J\\
\mu_{k}\left(J\right) &= \underset{\tilde{u}_k}{\text{min }} \mathbb{E}\left\{\left. \mu_{k+1}\left(\tilde{J}\right) \right| \mathcal{I}_k\right\}
\end{align}
with the cost function
\begin{align}
J &= x_{N+1}^T\bar{Q}x_{N+1} + \sum\limits_{k=k_0}^{N}\left(x_{k}^TQx_{k} + \sum_{p = \underline{d}}^{\overline{d}}\left[d_{k+p} = p\right]
\tilde{u}_{k}^{(p)^T}\!\! R \tilde{u}_{k}^{(p)} \right)\label{eq:5}
\end{align}
with $Q=Q^T\succeq 0$, $\bar{Q} = \bar{Q}^T \succeq 0$  and
$
\left[a = b\right] = \begin{cases}
1& \text{if } a=b\\ 0& \text{otherwise}
\end{cases}
$.

\subsubsection{Properties of the optimization problem}
Next, the following two properties of this optimization problem are shown:
\begin{itemize}
	\item Solving $\mu_{k_0}\left(J\right)$ yields the same values for $u_{k_0},...,u_{N}$ as solving $\mu_{k_0}\left(\tilde{J}\right)$ with
	\begin{align}
	\tilde{J} = x_{N+1}^T\bar{Q}x_{N+1} + \sum_{k=k_0}^{N}\left(x_{k}^TQx_{k} + u_{k}^TRu_{k}\right).\label{eq:2}
	\end{align}
	
	\item If $\mu_{k_0}\left(J\right)$ and 
	$\bar{\mu}_{k_0}\left(J\right) = \underset{\tilde{u}_{k_0},...,\tilde{u}_{N}}{\text{min }}\mathbb{E}\left\{\left. J \right|\mathcal{I}_{k_0}\right\}
	$
	exist, then
	\begin{align}\label{eq:1}
	\mu_{k_0}\left(J\right) \leq \bar{\mu}_{k_0}\left(J\right).
	\end{align}
\end{itemize}

In order to show \eqref{eq:1}, the compared optimization problems are written as
\begin{align}
\mu_{k_0}\left(J\right) &= 
\underset{\tilde{u}_k}{\text{min }} \mathbb{E}\left\{\left. \mu_{k_0+1}\left(J\right) \right| \mathcal{I}_{k_0}\right\}\\
&= 
\underset{\tilde{u}_k}{\text{min }} \mathbb{E}\left\{\left. 
\underset{\tilde{u}_{k_0+1}}{\text{min }} \mathbb{E}\left\{\left. 
\mu_{k_0+2}
\left(J\right) \right| \mathcal{I}_{k_0+1}\right\}
\left(J\right) \right| \mathcal{I}_{k_0}\right\}\\
&=
\underset{\tilde{u}_{k_0}}{\text{min }}\mathbb{E}\left\{\left.
\underset{\tilde{u}_{k_0+1}}{\text{min }}\mathbb{E}\left\{\left.
\ \hdots \
\underset{\tilde{u}_{N-1}}{\text{min }}\mathbb{E}\left\{\left.
\underset{\tilde{u}_{N}}{\text{min }}\mathbb{E}\left\{\left.
J
\right|\mathcal{I}_{N}\right\}
\right|\mathcal{I}_{N-1}\right\}
\ \hdots \
\right|\mathcal{I}_{k_0+1}\right\}
\right|\mathcal{I}_{k_0}\right\}
\end{align}
and
\begin{align}\bar{\mu}_{k_0}\left(J\right) &= \underset{\tilde{u}_{k_0},...,\tilde{u}_{N}}{\text{min }}\mathbb{E}\left\{\left. J \right|\mathcal{I}_{k_0}\right\}\\
&= 
\underset{\tilde{u}_{k_0}}{\text{min }}
\underset{\tilde{u}_{k_0+1}}{\text{min }}
\ \hdots \
\underset{\tilde{u}_{N-1}}{\text{min }}
\underset{\tilde{u}_{N}}{\text{min }}\mathbb{E}\left\{\left.
J
\right|\mathcal{I}_{k_0}\right\}\\
&= 
\underset{\tilde{u}_{k_0}}{\text{min }}
\underset{\tilde{u}_{k_0+1}}{\text{min }}
\ \hdots \
\underset{\tilde{u}_{N-1}}{\text{min }}
\underset{\tilde{u}_{N}}{\text{min }}\mathbb{E}\left\{\left.
\mathbb{E}\left\{\left.
\ \hdots \
\mathbb{E}\left\{\left.
\mathbb{E}\left\{\left.
J
\right|\mathcal{I}_{N}\right\}
\right|\mathcal{I}_{N-1}\right\}
\ \hdots \
\right|\mathcal{I}_{k_0+1}\right\}
\right|\mathcal{I}_{k_0}\right\}
\end{align}
since $\mathcal{I}_k \subseteq \mathcal{I}_{k+1}\ \forall k$ ("tower property" of the conditional expectation).\\

$\mu_{k_0}\left(J\right) \leq \bar{\mu}_{k_0}\left(J\right)$ if $\mu_{k_0}\left(J\right)$ and $\bar{\mu}_{k_0}\left(J\right)$ exist since
\begin{align}
\mathbb{E}\left\{\underset{a}{\text{min }}g(a,B)\right\}
\leq
\underset{a}{\text{min }}\mathbb{E}\left\{g(a,B)\right\}
\end{align}
for any discrete random variable $B$ and any function $g(a,B)$ such that above minima exist because
\begin{align}
\underset{a}{\text{min }}\mathbb{E}\left\{g(a,B)\right\}
&=
\underset{a}{\text{min}}\left(\sum_{b}\mathbb{P}(B=b)\, g(a,b)\right) =
\sum_{b}\mathbb{P}(B=b)\, f(a^*,b)
\end{align}
with 
\begin{align}
a^* = \underset{a}{\text{arg min}}\left(\sum_{b}\mathbb{P}(B=b)\, g(a,b)\right)
\end{align}
and
\begin{align}
\mathbb{E}\left\{\underset{a}{\text{min }}g(a,B)\right\}
&=
\mathbb{E}\left\{g(a^*_B,B)\right\} = 
\sum_{b}\mathbb{P}(B=b)\, g(a^*_b,b)
\end{align}
with 
\begin{align}
a^*_b = \underset{a}{\text{arg min }}g(a,b) \qquad &\Rightarrow \quad g(a^*_b,b) \leq g(a^*,b)\\
&\Rightarrow \quad 
\underbrace{\sum_{b}\mathbb{P}(B=b)\, g(a^*_b,b)}_{\mathbb{E}\left\{\underset{a}{\text{min }}g(a,B)\right\}} 
\ \leq\ 
\underbrace{\sum_{b}\mathbb{P}(B=b)\, g(a^*,b)}_{\underset{a}{\text{min }}\mathbb{E}\left\{g(a,B)\right\}}
.
\end{align}

In order show that solving $\mu_{k_0}\left(J\right)$ yields the same values for $u_{k_0},...,u_{N}$ as solving $\mu_{k_0}\left(\tilde{J}\right)$, the term quadratic in the actuating variable $u$ in
\begin{align}
\tilde{J} = x_{N+1}^T\bar{Q}x_{N+1} + \sum_{k=k_0}^{N}\left(x_{k}^TQx_{k} + u_{k}^TRu_{k}\right)\label{eq:6}
\end{align}
from \eqref{eq:2} is written as
\begin{align}
\sum_{k=k_0}^{N}u_{k}^TRu_{k} &= \sum_{k=k_0}^{N}\, \sum_{p=\underline{d}}^{\overline{d}}\left[d_k = p\right] u_{k}^TRu_{k}
= \sum_{p=\underline{d}}^{\overline{d}}\, \sum_{k=k_0}^{N}\left[d_k = p\right] u_{k}^TRu_{k} \label{eq:4}
\end{align}
where
$\sum_{p=\underline{d}}^{\overline{d}}\left[d_k = p\right] = 1
$
since $d_k \in \{\underline{d},...,\overline{d}\}$. Using \eqref{eq:2_1}, \eqref{eq:4} can be written as
\begin{align}
\sum_{k=k_0}^{N}u_{k}^TRu_{k} &= \sum_{p=\underline{d}}^{\overline{d}}\, \sum_{k=k_0}^{N}\left[d_k = p\right] u_{k}^TRu_{k}
= \sum_{p=\underline{d}}^{\overline{d}}\,\sum_{k=k_0}^{N}\left[d_k = p\right] \tilde{u}^{(d_k)^T}_{k-d_k}R\tilde{u}^{(d_k)}_{k-d_k}\\
&= \sum_{p=\underline{d}}^{\overline{d}}\,\sum_{k=k_0}^{N}\left[d_k = p\right] \tilde{u}^{(p)^T}_{k-p}R\tilde{u}^{(p)}_{k-p}
\end{align}
where replacing $k-p$ with $\tilde{k} = k-p$ so $k = \tilde{k}+p$ yields
\begin{align}
\sum_{k=k_0}^{N}u_{k}^TRu_{k} &= 
\sum_{p=\underline{d}}^{\overline{d}}\,\sum_{\tilde{k}=k_0-p}^{N-p}\left[d_{\tilde{k}+p} = p\right] \tilde{u}^{(p)^T}_{\tilde{k}}R\tilde{u}^{(p)}_{\tilde{k}}
=
U_1 + U_2 - U_3
\end{align}
with
\begin{align}
U_1 &= \label{eq:7}
\sum_{p = \underline{d}}^{\overline{d}}\ \sum_{\tilde{k} = k_0}^{N} \left[d_{\tilde{k}+p} = p\right] \tilde{u}_{\tilde{k}}^{(p)^T}R\tilde{u}_{\tilde{k}}^{(p)}\\
U_2 &= \label{eq:8}
\sum_{p = \underline{d}}^{\overline{d}}\ \sum_{\tilde{k} = k_0-p}^{k_0-1} \left[d_{\tilde{k}+p} = p\right] \tilde{u}_{\tilde{k}}^{(p)^T}R\tilde{u}_{\tilde{k}}^{(p)}\\
U_3 &= \label{eq:9}
\sum_{p = \underline{d}}^{\overline{d}}\ \sum_{\tilde{k} = N+1-p}^{N} \left[d_{\tilde{k}+p} = p\right] \tilde{u}_{\tilde{k}}^{(p)^T}R\tilde{u}_{\tilde{k}}^{(p)}
\end{align}
so \eqref{eq:6} can be written as
\begin{align}
\tilde{J} &= x_{N+1}^T\bar{Q}x_{N+1} + \sum_{k=k_0}^{N}x_{k}^TQx_{k} + U_1 + U_2 - U_3.
\end{align}

Replacing $\tilde{k}$ with $k$ in \eqref{eq:7} and using \eqref{eq:P_3} yields
\begin{align}
U_1
&=
\sum_{p = \underline{d}}^{\overline{d}}\, \sum_{k = k_0}^{N} \left[d_{k+p} = p\right] \tilde{u}_{k}^{(p)^T}R\tilde{u}_{k}^{(p)}
=
\sum_{k = k_0}^{N}\, \sum_{p = \underline{d}}^{\overline{d}}\  \left[d_{k+p} = p\right] \tilde{u}_{k}^{(p)^T}R\tilde{u}_{k}^{(p)}.
\end{align}

$U_2$ in \eqref{eq:8} only depends on $\tilde{u}_{k_0-\overline{d}},...,\tilde{u}_{k_0-1}$ but the optimization variables of $\mu_{k_0}\left(\tilde{J}\right)$ are $\tilde{u}_{k_0},...,\tilde{u}_{N}$. Therefore, solving $\mu_{k_0}\left(\tilde{J}\right)$ and $\mu_{k_0}\left(\tilde{J} - U_2\right)$ both yield the same optimal values for the controller outputs $\tilde{u}_{k_0},...,\tilde{u}_{N}$.

$U_3$ in \eqref{eq:9} only depends on $\tilde{u}^{(p)}_{\tilde{k}}$ with $\tilde{k} + p > N$ which only have an influence on the actuating variables $u_k$ with $k > N$ according to \eqref{eq:2_1}. Since $\tilde{J}$ is constant with respect to those actuating variables, solving
\begin{align}
\mu_{k_0}\left(\tilde{J} + \tilde{u}^{(p)^T}_{\tilde{k}} R \tilde{u}^{(p)}_{\tilde{k}}\right)
\qquad \text{ with } \tilde{k} + p > N \text{ and } k_0 \leq \tilde{k} \leq N
\end{align}
results in $\tilde{u}^{(p)}_{\tilde{k}} = 0$ because $R \succ 0$. Except for $\tilde{u}^{(p)}_{\tilde{k}}$, the optimal values for the controller outputs $\tilde{u}_{k_0},...,\tilde{u}_{N}$ are the same as for solving $\mu_{k_0}\left(\tilde{J}\right)$. Therefore, solving $\mu_{k_0}\left(\tilde{J}\right)$ and $\mu_{k_0}\left(\tilde{J} + U_3\right)$ results in the same actuating variables $u_{k_0},...,u_{N}$.

Summing up, solving $\mu_{k_0}\left(J - U_2 + U_3\right)$ results in the same actuating variables $u_{k_0},...,u_{N}$ as solving $\mu_{k_0}\left(J\right)$ where
\begin{align}
\tilde{J} - U_2 + U_3 &= x_{N+1}^T\bar{Q}x_{N+1} + \sum_{k=k_0}^{N}x_{k}^TQx_{k} + U_1\\
&= x_{N+1}^T\bar{Q}x_{N+1} + \sum_{k=k_0}^{N}x_{k}^TQx_{k} + \sum\limits_{k=k_0}^{N}\sum_{p = \underline{d}}^{\overline{d}}\left[d_{k+p} = p\right]
\tilde{u}_{k}^{(p)^T}\!\! R \tilde{u}_{k}^{(p)} = J
\end{align}
as defined in \eqref{eq:5}.

\subsubsection{Iteration law for solving the optimization problem}
The optimization problem to be solved is $\mu_{k_0}\left(J\right)$ with $\mu_{N+1}\left(J\right) = J$ and
\begin{align}
\mu_{k}\left(J\right) &= \underset{\tilde{u}_k}{\text{min }} \mathbb{E}\left\{\left. \mu_{k+1}\left(J\right) \right| \mathcal{I}_k\right\}
\end{align}
so
\begin{align}
\mu_{N}\left(J\right) &= 
\underset{\tilde{u}_{N}}{\text{min }} \mathbb{E}\left\{\left. \mu_{N+1}\left(J\right) \right| \mathcal{I}_{N}\right\}\\
&= \underset{\tilde{u}_{N}}{\text{min }} \mathbb{E}\left\{\left. J \right| \mathcal{I}_{N}\right\}\\
&= \underset{\tilde{u}_{N}}{\text{min }} \mathbb{E}\left\{\left. 
v_{N+1} + \sum_{k=k_0}^{N} f_k \right| \mathcal{I}_{N}\right\} \label{eq:10}
\end{align}
with
\begin{align}
f_k &= x_{k}^TQx_{k} + \sum_{p = \underline{d}}^{\overline{d}}\left[d_{k+p} = p\right]
\tilde{u}_{k}^{(p)^T}\!\! R \tilde{u}_{k}^{(p)}
\end{align}
and
\begin{align}
v_{N+1} &= \mathbb{E}\left\{\left. x_{N+1}^T\bar{Q}x_{N+1} \right| \mathcal{I}_{N+1}\right\}
\end{align}
since $\mathcal{I}_{N} \subseteq \mathcal{I}_{N+1}$ so
$
\mathbb{E}\left\{\left. \mathbb{E}\left\{\left. x_{N}^T\bar{Q}x_{N} \right| \mathcal{I}_{N+1}\right\} \right| \mathcal{I}_{N}\right\} = \mathbb{E}\left\{\left. x_{N}^T\bar{Q}x_{N} \right| \mathcal{I}_{N}\right\}
$. \eqref{eq:10} can be written as
\begin{align}
\mu_{N}\left(J\right)
&= \underset{\tilde{u}_{N}}{\text{min }} \mathbb{E}\left\{\left. 
v_{N+1} + \sum_{k=k_0}^{N} f_k \right| \mathcal{I}_{N}\right\}
=
\mathbb{E}\left\{\left. \sum_{k=k_0}^{N-1} f_k \right| \mathcal{I}_{N}\right\} + v_{N}
\end{align}
with
\begin{align}
v_{N} &= \underset{\tilde{u}_{N}}{\text{min }} \mathbb{E}\left\{\left. 
v_{N+1} +  f_{N} \right| \mathcal{I}_{N}\right\}
\end{align}
since $f_k$ with $k < N$ does not depend on $\tilde{u}_{N}$.\\

If $\mu_{k+1}\left(J\right)$ is given by
\begin{align}
\mu_{k+1}\left(J\right)
&=
\mathbb{E}\left\{\left. \sum_{\tilde{k}=k_0}^{k} f_{\tilde{k}} \right| \mathcal{I}_{k+1}\right\} + v_{k+1}
\end{align}
which applies for $k = N$, then
\begin{align}
\mu_{k}\left(J\right) &= \underset{\tilde{u}_k}{\text{min }} \mathbb{E}\left\{\left. \mu_{k+1}\left(J\right) \right| \mathcal{I}_k\right\}\\
&=
\underset{\tilde{u}_k}{\text{min }} \mathbb{E}\left\{\left.
\mathbb{E}\left\{\left. \sum_{\tilde{k}=k_0}^{k} f_{\tilde{k}} \right| \mathcal{I}_{k+1}\right\} + v_{k} \right| \mathcal{I}_k\right\}
=
\underset{\tilde{u}_k}{\text{min }} \mathbb{E}\left\{\left.
\sum_{\tilde{k}=k_0}^{k} f_{\tilde{k}} + v_{k+1} \right| \mathcal{I}_k\right\}\\
&=
\mathbb{E}\left\{\left.\sum_{\tilde{k}=k_0}^{k-1} f_{\tilde{k}} \right| \mathcal{I}_k\right\} + v_{k}
\end{align}
with
\begin{align}
v_{k} &= \underset{\tilde{u}_{k}}{\text{min }} \mathbb{E}\left\{\left. 
v_{k+1} +  f_{k} \right| \mathcal{I}_{k}\right\}
\end{align}
for $k \geq k_0$. Therefore, for $k = k_0$
\begin{align}
\mu_{k_0}\left(J\right) &= 
\mathbb{E}\left\{\left.\sum_{\tilde{k}=k_0}^{k_0-1} f_{\tilde{k}} \right| \mathcal{I}_{k_0}\right\} + v_{k_0}
= v_{k_0}.
\end{align}

Summing up, the optimization problem $\mu_{k_0}\left(J\right)$ can be solved by applying the iteration law
\begin{align}
v_{k} &= \underset{\tilde{u}_{k}}{\text{min }} \mathbb{E}\left\{\left. 
v_{k+1} + x_{k}^TQx_{k} + \sum_{p = \underline{d}}^{\overline{d}}\left[d_{k+p} = p\right]
\tilde{u}_{k}^{(p)^T}\!\! R \tilde{u}_{k}^{(p)} \right| \mathcal{I}_{k}\right\}\\
&= \underset{\tilde{u}_{k}}{\text{min }} \mathbb{E}\left\{\left. 
v_{k+1} + x_{k}^TQx_{k} + \tilde{u}_{k}^T \tilde{R}(d_{k+\underline{d}},\,...,\,d_{k+\overline{d}})\tilde{u}_{k} \right| \mathcal{I}_{k}\right\}\label{eq:20}
\end{align}
where
\begin{align}
&\tilde{R}(d_{k+\underline{d}},\,...,\,d_{k+\overline{d}}) =
\begin{bmatrix}
\left[d_{k+\overline{d}} = \overline{d}\right] R& & 0\\
& \ddots& \\
0& & \left[d_{k+\underline{d}} = \underline{d}\right] R
\end{bmatrix} \in\mathbb{R}^{\tilde{m}\times\tilde{m}}\label{eq:24}
\end{align}
for $k = N,...,k_0$ with the initial value
\begin{align}
v_{N+1} &= \mathbb{E}\left\{\left. x_{N+1}^T\bar{Q}x_{N+1} \right| \mathcal{I}_{N+1}\right\}.
\end{align}

\subsection{Solving the optimization problem}
First, an extended state vector $\hat{x}_k$ is defined. Then it is shown that
\begin{itemize}
	\item $v_{N+1}$ can be written as \begin{align}
	v_{N+1} &= \hat{x}_{N+1}^TK_{N+1}(r_{N+1}, \tilde{d}_{N+1})\hat{x}_{N+1}
	\end{align}
	with $K_{N+1}(r_{N+1}, \tilde{d}_{N+1}) = K_{N+1}^T(r_{N+1}, \tilde{d}_{N+1}) \succeq 0$.
	\item If $v_{k+1}$ for $k_0 \leq k \leq N$ can be written as
	\begin{align}
	v_{k+1} &= \hat{x}_{k+1}^TK_{k+1}(r_{k+1}, \tilde{d}_{k+1})\hat{x}_{k+1}
	\end{align}
	with $K_{k+1}(r_{k+1}, \tilde{d}_{k+1}) = K_{k+1}^T(r_{k+1}, \tilde{d}_{k+1}) \succeq 0$, then $v_{k}$ is given by
	\begin{align}
	v_{k} &= \hat{x}_{k}^TK_{k}(r_{k}, \tilde{d}_{k})\hat{x}_{k}
	\end{align}
	where again $K_{k}(r_{k}, \tilde{d}_{k}) = K_{k}^T(r_{k}, \tilde{d}_{k}) \succeq 0$ and the optimal controller output at time step $k$ is given by
	\begin{align}
	\tilde{u}_{k} &= -L_{k}(r_{k}, \tilde{d}_{k})\hat{x}_{k}.
	\end{align}
	\item $K_N(a,b)$, $K_{k}(a,b)$ and $L_{k}(a,b)$ for all $k_0 \leq k \leq N$ and all $a\in\{\underline{r},...,\overline{r}\}$, $b\in\{\underline{d},...,\overline{d}\}$ can be calculated offline in advance, i.e. without knowing the sequences $\mathbf{x}$, $\mathbf{u}$, $\mathbf{r}$ and $\mathbf{d}$ (or any partial sequences of these sequences).
\end{itemize}
In order to show above properties, the expected values
\begin{align}
E_1 &= \mathbb{E}\left\{\left. \tilde{u}_{k}^T \tilde{R}(d_{k+\underline{d}},\,...,\,d_{k+\overline{d}})\tilde{u}_{k} \right| \mathcal{I}_{k}\right\}\label{eq:21}\\
E_2 &= \mathbb{E}\left\{\left. 
x_{k}^TQx_{k} \right| \mathcal{I}_{k}\right\}\label{eq:22}\\
E_3 &= \mathbb{E}\left\{\left. 
v_{k+1} \right| \mathcal{I}_{k}\right\}\label{eq:23}
\end{align}
are calculated so \eqref{eq:20} can be written as
\begin{align}
v_{k} &= \underset{\tilde{u}_{k}}{\text{min }} \mathbb{E}\left\{\left. 
v_{k+1} + x_{k}^TQx_{k} + \tilde{u}_{k}^T \tilde{R}(d_{k+\underline{d}},\,...,\,d_{k+\overline{d}})\tilde{u}_{k} \right| \mathcal{I}_{k}\right\}\\
&= \underset{\tilde{u}_{k}}{\text{min }} \left(E_1 + E_2 + E_3\right)\label{eq:32}.
\end{align}

\subsubsection{Extended state vector}
The extended state vector $\hat{x}_k$ is given by
\begin{align}
\hat{x}_k &= \begin{bmatrix}
\tilde{x}_k\\ \hat{u}_k
\end{bmatrix}
\in\mathbb{R}^{n + \hat{m}}
\end{align}
with $\tilde{x}_k = x_{k-r_k}$ as defined in \eqref{eq:Ar3} and
\begin{align}
\hat{u}_k &= \begin{bmatrix}
\bar{u}_{k}(1)\\ \vdots\\ \bar{u}_{k}(\overline{d}+\overline{r})
\end{bmatrix} \in\mathbb{R}^{\hat{m}}
= \mathbb{R}^{\sum_{p = 1}^{\overline{d}+\overline{r}}\bar{m}_{p}} \label{eq:12}
\end{align}
where
\begin{align}
\bar{u}_{k}(p) &= \begin{bmatrix}
\tilde{u}_{k-p}^{(\overline{d})}\\ \vdots\\ \tilde{u}_{k-p}^{\left(\text{max}(\underline{d},\,p-\overline{r})\right)}
\end{bmatrix}
\in\mathbb{R}^{\left(1+\overline{d}-\text{max}(\underline{d},\,p-\overline{r})\right)m}
= \mathbb{R}^{\tilde{m} - \text{max}(0,\,p-\overline{r}-\underline{d})m}
= \mathbb{R}^{\bar{m}_{p}}.\label{ax:1}
\end{align}
For example,
\begin{align}
\bar{u}_{k}(0) &= \begin{bmatrix}
\tilde{u}_{k}^{(\overline{d})}\\ \vdots\\ \tilde{u}_{k}^{\left(\text{max}(\underline{d},\,-\overline{r})\right)}
\end{bmatrix}
=
\begin{bmatrix}
\tilde{u}_{k}^{(\overline{d})}\\ \vdots\\ \tilde{u}_{k}^{\left(\underline{d}\right)}
\end{bmatrix}
= 
\tilde{u}_{k}\label{eq:16}
\end{align}
and
\begin{align}
\bar{u}_{k}(p-1) &= \begin{bmatrix}
\tilde{u}_{k+1-p}^{(\overline{d})}\\ \vdots\\ \tilde{u}_{k+1-p}^{\left(\text{max}(\underline{d},\,p-1-\overline{r})\right)}
\end{bmatrix}
\in\mathbb{R}^{\bar{m}_{p-1}}
&
\bar{u}_{k+1}(p) = \begin{bmatrix}
\tilde{u}_{k+1-p}^{(\overline{d})}\\ \vdots\\ \tilde{u}_{k+1-p}^{\left(\text{max}(\underline{d},\,p-\overline{r})\right)}
\end{bmatrix}
\in\mathbb{R}^{\bar{m}_{p}}
\end{align}
so
\begin{align}
\bar{u}_{k+1}(p) &= T_p \bar{u}_{k}(p-1)\\
\bar{u}_{k+1}(1) &= T_0 \bar{u}_{k}(0) = T_0 \tilde{u}_{k}
\end{align}
with
\begin{align}
T_p &= \begin{cases} \label{eq:13}
I\in\mathbb{R}^{\tilde{m}\times\tilde{m}}& p \leq \overline{r}+\underline{d}\\
\begin{bmatrix}
I\in\mathbb{R}^{\bar{m}_{p}\times\bar{m}_{p}}& 0\in\mathbb{R}^{\bar{m}_{p}\times m}
\end{bmatrix}& \text{else}
\end{cases}.
\end{align}
From \eqref{eq:12} and \eqref{ax:1} follows that
\begin{align}
\hat{u}_{k+1} &= \begin{bmatrix}
\bar{u}_{k+1}(1)\\ \vdots\\ \bar{u}_{k+1}(\overline{d}+\overline{r})
\end{bmatrix} = \begin{bmatrix}
\bar{u}_{k+1}(1)\\ \vdots\\ \bar{u}_{k+1}(\overline{d}+\overline{r})\\ \bar{u}_{k+1}(\overline{d}+\overline{r}+1)
\end{bmatrix}
\end{align}
since $\bar{u}_{k+1}(\overline{d}+\overline{r}+1)\in\mathbb{R}^0$. Using \eqref{eq:13}, this can be written as
\begin{align}
\hat{u}_{k+1} &= 
\begin{bmatrix}
T_{0}\tilde{u}_{k}\\ T_{1}\bar{u}_{k}(0)\\ \vdots\\ T_{\overline{d}+\overline{r}}\bar{u}_{k}(\overline{d}+\overline{r})
\end{bmatrix}
= 
\begin{bmatrix}
T_0& 0\\ 0& \bar{T}
\end{bmatrix}\begin{bmatrix}
\tilde{u}_k\\ \hat{u}_k
\end{bmatrix}
= \hat{T}\hat{u}_k + \tilde{T}\tilde{u}_k \label{eq:19}
\end{align}
with
\begin{align}
\bar{T} &= \begin{bmatrix}
T_1& & 0\\
& \ddots& \\
0& & T_{\overline{d}+\overline{r}}
\end{bmatrix}
\in\mathbb{R}^{(\hat{m}-\tilde{m})\times\hat{m}}
\end{align}
and
\begin{align}
\hat{T} &= \begin{bmatrix}
0\in\mathbb{R}^{\tilde{m} \times \hat{m}}\\ \bar{T}
\end{bmatrix}
\in\mathbb{R}^{\hat{m}\times\hat{m}}\\
\tilde{T} &= \begin{bmatrix}
T_0\\ 0\in\mathbb{R}^{(\hat{m}-\tilde{m})\times \tilde{m}}
\end{bmatrix}
\in\mathbb{R}^{\hat{m}\times\tilde{m}}.
\end{align}
The state vector $\tilde{x}_{k+1} = x_{k+1-r_{k+1}}$ is given by
\begin{align}
\tilde{x}_{k+1} &= A^{(1+r_k-r_{k+1})}\tilde{x}_k + \sum_{i=0}^{r_k-r_{k+1}}A^iBu_{k-r_{k+1}-i}\\
&= A^{(1+r_k-r_{k+1})}\tilde{x}_k + \sum_{i=r_{k+1}}^{r_k}A^{(i-r_{k+1})}Bu_{k-i}\label{eq:17}
\end{align}
where
\begin{align}
u_{k-i} &= \tilde{u}_{k-i-d_{k-i}}^{(d_{k-i})}
\end{align}
which is contained in $\bar{u}_{k}(i+d_{k-i})$ if 
\begin{align}
d_{k-i} \geq \text{max}(\underline{d},\,i+d_{k-i}-\overline{r})
\quad\Leftrightarrow\quad
d_{k-i} \geq i+d_{k-i}-\overline{r}
\quad\Leftrightarrow\quad
i\leq\overline{r}
\end{align}
so
\begin{align}
u_{k-i} &= \check{\mathbb{I}}_i(d_{k-i}) \bar{u}_{k}(i+d_{k-i}) \label{eq:14}
\end{align}
with
\begin{align}
\check{\mathbb{I}}_i(d_{k-i}) &= \begin{bmatrix}
0\in\mathbb{R}^{m \times (\overline{d}-d_{k-i})m}& I\in\mathbb{R}^{m \times m} & 0\in\mathbb{R}^{m \times \bar{m}_{i+d_{k-i}} - (1+\overline{d}-d_{k-i})m}
\end{bmatrix}
\in\mathbb{R}^{m\times \bar{m}_{i+d_{k-i}}}.
\end{align}
Due to \eqref{eq:12}
\begin{align}
\bar{u}_{k}(p) &= \bar{\mathbb{I}}_p \hat{u}_k \qquad \text{for } 1\leq p\leq \overline{d}+\overline{r} \label{eq:15}
\end{align}
with
\begin{align}
\bar{\mathbb{I}}_p &= \begin{cases}
\begin{bmatrix}
0\in\mathbb{R}^{\bar{m}_p \times \sum_{i=1}^{p-1}\bar{m}_i}& I\in\mathbb{R}^{\bar{m}_p \times \bar{m}_p} & 0\in\mathbb{R}^{\bar{m}_p \times \hat{m}-\sum_{i=1}^{p}\bar{m}_i}
\end{bmatrix}& 1\leq p\leq \overline{d}+\overline{r}\\
0 \in\mathbb{R}^{\bar{m}_p\times \hat{m}}& \text{else}
\end{cases}\\
&\in\mathbb{R}^{\bar{m}_p\times \hat{m}}.
\end{align}
Using \eqref{eq:16} and \eqref{eq:15}, \eqref{eq:14} can be written as
\begin{align}
u_{k-i} &= 
\check{\mathbb{I}}_i(d_{k-i}) \bar{u}_{k}(i+d_{k-i})\\
&= \begin{cases}
\check{\mathbb{I}}_i(d_{k-i})
\bar{\mathbb{I}}_{i+d_{k-i}} \hat{u}_k& 1\leq i \leq \overline{r}\\
\check{\mathbb{I}}_i(d_{k})
\bar{\mathbb{I}}_{d_k} \hat{u}_k & i=0,\,d_k \geq 1\\
\tilde{u}_k^{(0)} & i=0,\,d_k=0
\end{cases}\\
&= \begin{cases}
\hat{\mathbb{I}}_i(d_{k-i}) \hat{u}_k & 1\leq i \leq \overline{r}\\
\hat{\mathbb{I}}_0(d_k) \hat{u}_k + \tilde{\mathbb{I}}(d_k)\tilde{u}_k & i=0
\end{cases}\label{eq:18}
\end{align}
with
\begin{align}
\hat{\mathbb{I}}_i(d_{k-i}) &= \check{\mathbb{I}}_i(d_{k-i}) \bar{\mathbb{I}}_{i+d_{k-i}}
\in\mathbb{R}^{m\times \hat{m}}\\
\tilde{\mathbb{I}}(d_{k}) &= \left[d_{k} = 0\right] \begin{bmatrix}
0\in\mathbb{R}^{m\times (\overline{d}-\underline{d})m}& I\in\mathbb{R}^{m\times m}
\end{bmatrix}
\in\mathbb{R}^{m\times \tilde{m}}.
\end{align}
Inserting \eqref{eq:18} in
$\sum_{i=r_{k+1}}^{r_k}A^{(i-r_{k+1})}Bu_{k-i}
$
from \eqref{eq:17} yields
\begin{align}
\sum_{i=r_{k+1}}^{r_k}A^{(i-r_{k+1})}Bu_{k-i} 
&= \begin{cases}
\sum_{i=1}^{r_k}A^{i}B\,u_{k-i} + B\,u_{k}& r_{k+1} = 0\\
\sum_{i=r_{k+1}}^{r_k}A^{(i-r_{k+1})}B\,u_{k-i}& \text{else}
\end{cases}\\
&= \begin{cases}
\sum_{i=1}^{r_k}A^{i}B\,\hat{\mathbb{I}}_i(d_{k-i}) \hat{u}_k + B\left(\hat{\mathbb{I}}_0(d_{k})  \hat{u}_k + \tilde{\mathbb{I}}(d_{k})\tilde{u}_k\right)& r_{k+1} = 0\\
\sum_{i=r_{k+1}}^{r_k}A^{(i-r_{k+1})}B\,\hat{\mathbb{I}}_i(d_{k-i}) \hat{u}_k& \text{else}
\end{cases}\\
&= \begin{cases}
\sum_{i=0}^{r_k}A^{i}B\,\hat{\mathbb{I}}_i(d_{k-i}) \hat{u}_k + B\,\tilde{\mathbb{I}}(d_{k})\tilde{u}_k& r_{k+1} = 0\\
\sum_{i=r_{k+1}}^{r_k}A^{(i-r_{k+1})}B\,\hat{\mathbb{I}}_i(d_{k-i}) \hat{u}_k& \text{else}
\end{cases}\\
&=
\sum_{i=r_{k+1}}^{r_k}A^{(i-r_{k+1})}B\,\hat{\mathbb{I}}_i(d_{k-i}) \hat{u}_k + [r_{k+1}=0]B\,\tilde{\mathbb{I}}(d_{k})\tilde{u}_k
\end{align}
so
\begin{align}
\tilde{x}_{k+1} &= 
A^{(1+r_k-r_{k+1})}\tilde{x}_k + \sum_{i=r_{k+1}}^{r_k}A^{(i-r_{k+1})}B\,\hat{\mathbb{I}}_i(d_{k-i}) \hat{u}_k + [r_{k+1}=0]B\,\tilde{\mathbb{I}}(d_{k})\tilde{u}_k
\end{align}
and with \eqref{eq:19}
\begin{align}
\begin{bmatrix}
\tilde{x}_{k+1}\\ \hat{u}_{k+1}
\end{bmatrix}
&= 
\begin{bmatrix}
A^{(1+r_k-r_{k+1})}& \sum_{i=r_{k+1}}^{r_k}A^{(i-r_{k+1})}B\,\hat{\mathbb{I}}_i(d_{k-i})\\
0\in\mathbb{R}^{\bar{m}\times n}& \hat{T}
\end{bmatrix}
\begin{bmatrix}
\tilde{x}_k\\ \hat{u}_k
\end{bmatrix}
+
\begin{bmatrix}
[r_{k+1}=0]B\,\tilde{\mathbb{I}}(d_{k})\\ \tilde{T}
\end{bmatrix}\tilde{u}_k
\end{align}
and therefore
\begin{align}
\hat{x}_{k+1} &= 	
\tilde{A}(r_k,r_{k+1})\hat{x}_k + \sum\limits_{i=r_{k+1}}^{r_k}\bar{A}_{i}(r_{k+1},d_{k-i})\hat{x}_k + \tilde{B}(r_{k+1},d_k)\tilde{u}_k \label{eq:31}
\end{align}
with
\begin{align}
\tilde{A}(r_k,r_{k+1}) &= \begin{bmatrix}
A^{(1+r_k-r_{k+1})}& 0\in\mathbb{R}^{n \times \hat{m}}\\
0\in\mathbb{R}^{\hat{m}\times n}& \hat{T}
\end{bmatrix}\in\mathbb{R}^{(n+\hat{m})\times (n+\hat{m})}\\
\bar{A}_{i}(r_{k+1},d_{k-i}) &= \begin{bmatrix}
0\in\mathbb{R}^{n\times n}& A^{(i-r_{k+1})}B\,\hat{\mathbb{I}}_i(d_{k-i})\\
0\in\mathbb{R}^{\hat{m}\times n}& 0\in\mathbb{R}^{\hat{m}\times \hat{m}}
\end{bmatrix}\in\mathbb{R}^{(n+\hat{m})\times (n+\hat{m})}\\
\tilde{B}(r_{k+1},d_k) &= \begin{bmatrix}
[r_{k+1}=0]B\,\tilde{\mathbb{I}}(d_{k})\\ \tilde{T}
\end{bmatrix}\in\mathbb{R}^{(n+\hat{m})\times \tilde{m}}.
\end{align}
\eqref{eq:18} can also be used to write the state vector $x_k$ as
\begin{align}
x_k &= A^{r_k}\tilde{x}_k + \sum_{i=1}^{r_k}A^{(i-1)}Bu_{k-i}\\
&= A^{r_k}\tilde{x}_k + \sum_{i=1}^{r_k}A^{(i-1)}B\,\hat{\mathbb{I}}_i(d_{k-i})\hat{u}_k\label{eq:26}.
\end{align}

\subsubsection{Expected value $E_1$}
Using \eqref{eq:24}, the expected value $E_1$ from \eqref{eq:21} can be written as
\begin{align}
E_1 &= \mathbb{E}\left\{\left. \tilde{u}_{k}^T \tilde{R}(\mathcal{D}_{k+\underline{d}},\,...,\,\mathcal{D}_{k+\overline{d}})\tilde{u}_{k} \right| \mathcal{I}_{k}\right\}\\
&=
\tilde{u}_{k}^T \mathbb{E}\left\{\left.  \tilde{R}(\mathcal{D}_{k+\underline{d}},\,...,\,\mathcal{D}_{k+\overline{d}}) \right| \mathcal{I}_{k}\right\}\tilde{u}_{k}\\
&=\tilde{u}_{k}^T \hat{R}(r_k,\tilde{d}_{k}) \tilde{u}_{k}\label{eq:33}
\end{align}
with
\begin{align}
\hat{R}(r_k,\tilde{d}_{k}) &=
\mathbb{E}\left\{\tilde{R}(\mathcal{D}_{k+\underline{d}},\,...,\,\mathcal{D}_{k+\overline{d}})\big|\mathcal{I}_{k}\right\}\\
&=
\mathbb{E}\left\{\left.
\begin{bmatrix}
\left[\mathcal{D}_{k+\overline{d}} = \overline{d}\right] R& & 0\\
& \ddots& \\
0& & \left[\mathcal{D}_{k+\underline{d}} = \underline{d}\right] R
\end{bmatrix}
\right|\mathcal{I}_{k}\right\}\\
&=
\begin{bmatrix}
\mathbb{P}\left(\mathcal{D}_{k+\overline{d}}=\overline{d} \,|\, \mathcal{I}_{k}\right)R& & 0\\
& \ddots& \\
0& & \mathbb{P}\left(\mathcal{D}_{k+\underline{d}}=\underline{d} \,|\, \mathcal{I}_{k}\right)R
\end{bmatrix}\label{eq:25}
\end{align}
where
\begin{align}
\mathbb{P}\left(\mathcal{D}_{k+p}=p \,|\, \mathcal{I}_{k}\right) \qquad \text{with } p \in \left\{\underline{d},...,\overline{d}\right\}
\end{align}
can be written as
\begin{align}
\mathbb{P}\left(\mathcal{D}_{k+p}=p \,|\, \mathcal{I}_{k}\right)
&=
\mathbb{P}\left(\mathcal{D}_{k+p}=p \,|\, \mathcal{D}_{k-1-r_k}= \tilde{d}_{k}\right)\\
&=
\Phi_{(1+r_k+p)}(\tilde{d}_k,p)
\end{align}
using \eqref{eq2_11}, since $d_{k-1-r_k} = \tilde{d}_{k}$ is the most recent available value in $\mathbf{d}$ and $k+p \geq k-1-r_k$. Therefore,
\eqref{eq:25} can be written as
\begin{align}
\hat{R}(r_k,\tilde{d}_{k})&=
\begin{bmatrix}
\Phi_{(1+r_k+\overline{d})}(\tilde{d}_k,\overline{d})R& & 0\\
& \ddots& \\
0& & \Phi_{(1+r_k+\underline{d})}(\tilde{d}_k,\underline{d})R
\end{bmatrix} \in\mathbb{R}^{\tilde{m}\times\tilde{m}}
\end{align}
where $\hat{R}(r_k,\tilde{d}_{k}) = \hat{R}^T(r_k,\tilde{d}_{k}) \succ 0$ since $R = R^T \succ 0$ and $\Phi_{(1+r_k+p)}(\tilde{d}_k,p) > 0$ for $p \in \left\{\underline{d},...,\overline{d}\right\}$ due to \eqref{eq3_9}.

\subsubsection{Expected value $E_2$}
Using \eqref{eq:26}, the term $x_{k}^TQx_{k}$ in $E_2$ from \eqref{eq:22} can be written as
\begin{align}
x_k^TQx_k &= \left(A^{r_k}\tilde{x}_k + \sum_{i=1}^{r_k}A^{(i-1)}B\,\hat{\mathbb{I}}_i(d_{k-i})\hat{u}_k\right)^TQ\left(A^{r_k}\tilde{x}_k + \sum_{i=1}^{r_k}A^{(i-1)}B\,\hat{\mathbb{I}}_i(d_{k-i})\hat{u}_k\right)\\
&= \tilde{x}_k^T A^{r_k^T} Q A^{r_k}\tilde{x}_k
+ 2\tilde{x}_k^T A^{r_k^T} Q \sum_{i=1}^{r_k}A^{(i-1)}B\,\hat{\mathbb{I}}_i(d_{k-i})\hat{u}_k\\
&\qquad + \sum_{i=1}^{r_k}\hat{u}^T_k\,\hat{\mathbb{I}}^T_i(d_{k-i}) B^TA^{(i-1)^T} Q \sum_{j=1}^{r_k}A^{(j-1)}B\,\hat{\mathbb{I}}_j(d_{k-j})\hat{u}_k\nonumber\\
&= \tilde{x}_k^T\left( A^{r_k^T} Q A^{r_k}\right)\tilde{x}_k
+ 2\tilde{x}_k^T\left(\sum_{i=1}^{r_k} A^{r_k^T} Q A^{(i-1)}B\,\hat{\mathbb{I}}_i(d_{k-i})\right)\hat{u}_k\\
&\qquad + \hat{u}^T_k\left(\sum_{i=1}^{r_k}\sum_{j=1}^{r_k}\,\hat{\mathbb{I}}^T_i(d_{k-i}) B^TA^{(i-1)^T} Q A^{(j-1)}B\,\hat{\mathbb{I}}_j(d_{k-j})\right)\hat{u}_k\nonumber\\
&= \hat{x}_k^T \tilde{Q}(r_k,d_{k-r_k},...,d_{k-1}) \hat{x}_k \label{eq:27}
\end{align}
with
\begin{align}
\tilde{Q}(r_k,d_{k-r_k},...,d_{k-1}) &= \begin{bmatrix}
\tilde{Q}_{11}(r_k)& \tilde{Q}_{12}(r_k,d_{k-r_k},...,d_{k-1})\\ \tilde{Q}^T_{12}(r_k,d_{k-r_k},...,d_{k-1})& \tilde{Q}_{22}(r_k,d_{k-r_k},...,d_{k-1})
\end{bmatrix}
\in\mathbb{R}^{(n+\hat{m})\times (n+\hat{m})}
\end{align}
where
\begin{align}
\tilde{Q}_{11}(r_k) &= A^{r_k^T} Q A^{r_k} \in\mathbb{R}^{n\times n}\\
\tilde{Q}_{12}(r_k,d_{k-r_k},...,d_{k-1}) &= \sum_{i=1}^{r_k} A^{r_k^T} Q A^{(i-1)}B\,\hat{\mathbb{I}}_i(d_{k-i}) \in\mathbb{R}^{n\times \hat{m}}\\
\tilde{Q}_{22}(r_k,d_{k-r_k},...,d_{k-1}) &= \sum_{i=1}^{r_k}\sum_{j=1}^{r_k}\,\hat{\mathbb{I}}^T_i(d_{k-i}) B^TA^{(i-1)^T} Q A^{(j-1)}B\,\hat{\mathbb{I}}_j(d_{k-j}) \in\mathbb{R}^{\hat{m}\times \hat{m}}
\end{align}
and $\tilde{Q}(r_k,d_{k-r_k},...,d_{k-1}) = \tilde{Q}^T(r_k,d_{k-r_k},...,d_{k-1}) \succeq 0$ since $\hat{x}_k^T \tilde{Q}(r_k,d_{k-r_k},...,d_{k-1}) \hat{x}_k = x_k^TQx_k$ and $x_k^TQx_k \geq 0$ for any $\hat{x}_k$.

Using \eqref{eq:27}, the expected value $E_2$ from \eqref{eq:22} can be written as
\begin{align}
E_2 &= \mathbb{E}\left\{\left.x_k^TQx_k \,\right|\, \mathcal{I}_k\right\}\\
&= 
\mathbb{E}\left\{\left.\hat{x}_k^T \tilde{Q}(r_k,\mathcal{D}_{k-r_k},...,\mathcal{D}_{k-1}) \hat{x}_k \,\right|\, \mathcal{I}_k\right\}
= 
\hat{x}_k^T \mathbb{E}\left\{\left.\tilde{Q}(r_k,\mathcal{D}_{k-r_k},...,\mathcal{D}_{k-1}) \,\right|\, \mathcal{I}_k\right\} \hat{x}_k\\
&=
\hat{x}_k^T \hat{Q}(r_k,\tilde{d}_k)\hat{x}_k \label{eq:29}
\end{align}
with 
\begin{align}
\hat{Q}(r_k,\tilde{d}_k) &= \begin{bmatrix}
\hat{Q}_{11}(r_k)& \hat{Q}_{12}(r_k,\tilde{d}_k)\\ \hat{Q}^T_{12}(r_k,\tilde{d}_k)& \hat{Q}_{22}(r_k,\tilde{d}_k)
\end{bmatrix}
\end{align}
where
\begin{align}
\hat{Q}_{11}(r_k) &= \mathbb{E}\left\{\left.\tilde{Q}_{11}(r_k)\,\right|\, \mathcal{I}_k\right\}  = \tilde{Q}_{11}(r_k) = A^{r_k^T} Q A^{r_k},
\end{align}
\begin{align}
\hat{Q}_{12}(r_k,\tilde{d}_k) &= \mathbb{E}\left\{\left.\tilde{Q}_{12}(r_k,\mathcal{D}_{k-r_k},...,\mathcal{D}_{k-1})\,\right|\, \mathcal{I}_k\right\}\\
&= \mathbb{E}\left\{\left.\sum_{i=1}^{r_k} A^{r_k^T} Q A^{(i-1)}B\,\hat{\mathbb{I}}_i(\mathcal{D}_{k-i})\,\right|\, \mathcal{I}_k\right\}\\
&= \sum_{i=1}^{r_k}\left( A^{r_k^T} Q A^{(i-1)}B\,\mathbb{E}\left\{\left.\hat{\mathbb{I}}_i(\mathcal{D}_{k-i})\,\right|\, \mathcal{I}_k\right\}\right)\\
&=
\sum_{i=1}^{r_k}\left( A^{r_k^T} Q A^{(i-1)}B\,\sum_{\delta = \underline{d}}^{\overline{d}}\mathbb{P}\left(\left.\mathcal{D}_{k-i} = \delta\,\right.|\,\mathcal{I}_k\right)\hat{\mathbb{I}}_i(\delta)\right) \label{eq:28}
\end{align}
where
\begin{align}
\mathbb{P}\left(\left.\mathcal{D}_{k-i} = \delta\,\right.|\,\mathcal{I}_k\right) \qquad \text{with } i \leq r_k
\end{align}
can be written as
\begin{align}
\mathbb{P}\left(\left.\mathcal{D}_{k-i} = \delta\,\right.|\,\mathcal{I}_k\right) &=
\mathbb{P}\left(\left.\mathcal{D}_{k-i} = \delta\,\right.|\,\mathcal{D}_{k-1-r_k}=\tilde{d}_k\right)
= \Phi_{(1+i+r_k)}(\tilde{d}_k,\delta)
\end{align}
using \eqref{eq2_11}, since $d_{k-1-r_k} = \tilde{d}_{k}$ is the most recent available value in $\mathbf{d}$ and $k-i \geq k-1-r_k$. Therefore, \eqref{eq:28} can be written as
\begin{align}
\hat{Q}_{12}(r_k,\tilde{d}_k)&=
\sum_{i=1}^{r_k} A^{r_k^T} Q A^{(i-1)}B\,\sum_{\delta = \underline{d}}^{\overline{d}}\Phi_{(1+i+r_k)}(\tilde{d}_k,\delta)\,\hat{\mathbb{I}}_i(\delta).
\end{align}
\begin{align}
\hat{Q}_{22}(r_k,\tilde{d}_k) &=\mathbb{E}\left\{\left.\tilde{Q}_{22}(r_k,\mathcal{D}_{k-r_k},...,\mathcal{D}_{k-1})\,\right|\, \mathcal{I}_k\right\}\\
&=\mathbb{E}\left\{\left.\sum_{i=1}^{r_k}\sum_{j=1}^{r_k}\,\hat{\mathbb{I}}^T_i(\mathcal{D}_{k-i}) B^TA^{(i-1)^T} Q A^{(j-1)}B\,\hat{\mathbb{I}}_j(\mathcal{D}_{k-j})\,\right|\, \mathcal{I}_k\right\}\\
&=\sum_{i=1}^{r_k}\sum_{j=1}^{r_k}\,\mathbb{E}\left\{\left.\hat{\mathbb{I}}^T_i(\mathcal{D}_{k-i}) B^TA^{(i-1)^T} Q A^{(j-1)}B\,\hat{\mathbb{I}}_j(\mathcal{D}_{k-j})\,\right|\, \mathcal{I}_k\right\}\\
&=\sum_{i=1}^{r_k}\sum_{j=1}^{r_k}\sum_{\delta_1 = \underline{d}}^{\overline{d}}\sum_{\delta_2 = \underline{d}}^{\overline{d}}\mathcal{P}(r_k,\tilde{d}_k,i,j,\delta_1,\delta_2)\,\hat{\mathbb{I}}^T_i(\delta_1) B^TA^{(i-1)^T} Q A^{(j-1)}B\,\hat{\mathbb{I}}_j(\delta_2)\\
&=\sum_{i=1}^{r_k}\sum_{\delta_1 = \underline{d}}^{\overline{d}}\,\hat{\mathbb{I}}^T_i(\delta_1) B^TA^{(i-1)^T} Q \sum_{j=1}^{r_k} A^{(j-1)}B \sum_{\delta_2 = \underline{d}}^{\overline{d}}
\mathcal{P}(r_k,\tilde{d}_k,i,j,\delta_1,\delta_2)\,\hat{\mathbb{I}}_j(\delta_2)
\end{align}
where
\begin{align}
&\mathcal{P}(r_k,\tilde{d}_k,i,j,\delta_1,\delta_2) = 
\mathbb{P}\left(\left.\mathcal{D}_{k-i} = \delta_1\,\right.|\,\mathcal{I}_k\right)
\mathbb{P}\left(\left.\mathcal{D}_{k-j} = \delta_2\,\right.|\,\mathcal{I}_k, \mathcal{D}_{k-i} = \delta_1 \right)\\ &\qquad= 
\mathbb{P}\left(\left.\mathcal{D}_{k-i} = \delta_1\,\right.|\,\mathcal{D}_{k-1-r_k}=\tilde{d}_k\right)
\mathbb{P}\left(\left.\mathcal{D}_{k-j} = \delta_2\,\right.|\,\mathcal{D}_{k-1-r_k}=\tilde{d}_k, \mathcal{D}_{k-i} = \delta_1 \right)\\ &\qquad= 
\mathbb{P}\left(\left.\mathcal{D}_{k-j} = \delta_2\,\right.|\,\mathcal{D}_{k-1-r_k}=\tilde{d}_k\right)
\mathbb{P}\left(\left.\mathcal{D}_{k-i} = \delta_1\,\right.|\,\mathcal{D}_{k-1-r_k}=\tilde{d}_k, \mathcal{D}_{k-j} = \delta_2 \right)\end{align}
which can be written as
\begin{align} 
&\mathcal{P}(r_k,\tilde{d}_k,i,j,\delta_1,\delta_2)\nonumber\\
&\qquad= 
\begin{cases}
\mathbb{P}\left(\left.\mathcal{D}_{k-i} = \delta_1\,\right.|\,\mathcal{D}_{k-1-r_k}=\tilde{d}_k\right)
\mathbb{P}\left(\left.\mathcal{D}_{k-j} = \delta_2\,\right.|\,\mathcal{D}_{k-i} = \delta_1 \right)& i\geq j\\
\mathbb{P}\left(\left.\mathcal{D}_{k-j} = \delta_2\,\right.|\,\mathcal{D}_{k-1-r_k}=\tilde{d}_k\right)
\mathbb{P}\left(\left.\mathcal{D}_{k-i} = \delta_1\,\right.|\, \mathcal{D}_{k-j} = \delta_2 \right)&
i\leq j
\end{cases}\\ &\qquad= 
\begin{cases}
\Phi_{(1+i+r_k)}(\tilde{d}_k,\delta_1)\,\Phi_{(i-j)}(\delta_1,\delta_2)& i\geq j\\
\Phi_{(1+j+r_k)}(\tilde{d}_k,\delta_2)\,\Phi_{(j-i)}(\delta_2,\delta_1)& i\leq j
\end{cases}\\ &\qquad= 
\begin{cases}
\Phi_{(1+i+r_k)}(\tilde{d}_k,\delta_1)\,\Phi_{(i-j)}(\delta_1,\delta_2)& i>j\\
\Phi_{(1+j+r_k)}(\tilde{d}_k,\delta_2)\,\Phi_{(j-i)}(\delta_2,\delta_1)& i<j\\
\Phi_{(1+i+r_k)}(\tilde{d}_k,\delta_1)& i=j,\delta_1=\delta_2\\
0& i=j,\delta_1\neq\delta_2
\end{cases}
\end{align}
using \eqref{eq2_11}, since \begin{itemize}
	\item $d_{k-1-r_k} = \tilde{d}_{k}$ is the most recent available value in $\mathbf{d}$,
	\item $k-j \geq k-i \geq k-1-r_k$ for $i\geq j$,
	\item and $k-i \geq k-j \geq k-1-r_k$ for $i\leq j$.
\end{itemize}
Since $\hat{Q}_{11}(r_k) = \hat{Q}^T_{11}(r_k)$ and $\hat{Q}_{22}(r_k,\tilde{d}_k) = \hat{Q}_{22}^T(r_k,\tilde{d}_k)$ also $\hat{Q}(r_k,\tilde{d}_k) = \hat{Q}^T(r_k,\tilde{d}_k)$.

\subsubsection{Expected value $E_3$}
It is assumed that $v_{k+1}$ from \eqref{eq:23} can be written as 	
\begin{align}
v_{k+1} &= \hat{x}_{k+1}^TK_{k+1}(r_{k+1}, \tilde{d}_{k+1})\hat{x}_{k+1}\label{eq:30}
\end{align}
with $K_{k+1}(r_{k+1}, \tilde{d}_{k+1}) = K_{k+1}^T(r_{k+1}, \tilde{d}_{k+1}) \succeq 0$. This is fulfilled for $k=N$ since
\begin{align}
v_{N+1} &= 
\mathbb{E}\left\{\left.x_{N+1}^T\bar{Q}x_{N+1}\right|\mathcal{I}_{N+1}\right\}\\
&=
\hat{x}_{N+1}^T K_{N+1}(r_{N+1},\tilde{d}_{N+1})\hat{x}_{N+1}
\end{align}
where $K_{N+1}(r_{N+1},\tilde{d}_{N+1}) = K_{N+1}^T(r_{N+1},\tilde{d}_{N+1}) \succeq 0$ can be obtained using the results from \eqref{eq:29} and is given by
\begin{align}
K_{N+1}(r_{N+1},\tilde{d}_{N+1}) &= \begin{bmatrix}
\hat{\bar{Q}}_{11}(r_{N+1})& \hat{\bar{Q}}_{12}(r_{N+1},\tilde{d}_{N+1})\\ \hat{\bar{Q}}^T_{12}(r_{N+1},\tilde{d}_{N+1})& \hat{\bar{Q}}_{22}(r_{N+1},\tilde{d}_{N+1})
\end{bmatrix}
\end{align}
with
\begin{align}
\hat{\bar{Q}}_{11}(r_{N+1}) &= A^{r_{N+1}^T} \bar{Q} A^{r_{N+1}}\\
\hat{\bar{Q}}_{12}(r_{N+1},\tilde{d}_{N+1}) &=
\sum_{i=1}^{r_{N+1}} A^{r_{N+1}^T} \bar{Q} A^{(i-1)}B\,\sum_{\delta = \underline{d}}^{\overline{d}}\Phi_{(1+i+r_{N+1})}(\tilde{d}_{N+1},\delta)\,\hat{\mathbb{I}}_i(\delta)\\
\hat{\bar{Q}}_{22}(r_{N+1},\tilde{d}_{N+1})
&=\sum_{i=1}^{r_{N+1}}\sum_{\delta_1 = \underline{d}}^{\overline{d}}\,\hat{\mathbb{I}}^T_i(\delta_1) B^TA^{(i-1)^T} \bar{Q} \sum_{j=1}^{r_{N+1}} A^{(j-1)}B \sum_{\delta_2 = \underline{d}}^{\overline{d}}
\mathcal{P}(r_{N+1},\tilde{d}_{N+1},i,j,\delta_1,\delta_2)\,\hat{\mathbb{I}}_j(\delta_2).
\end{align}

Using \eqref{eq:31}, \eqref{eq:30} can be written as
\begin{align}
v_{k+1} &= \hat{x}_{k+1}^TK_{k+1}(r_{k+1},\tilde{d}_{k+1})\hat{x}_{k+1}\\
&=
\tilde{u}_k^T\tilde{B}^T(r_{k+1},d_k) K_{k+1}(r_{k+1},\tilde{d}_{k+1}) \tilde{B}(r_{k+1},d_k)\tilde{u}_k\\
&+ 2\,
\tilde{u}_k^T\tilde{B}^T(r_{k+1},d_k) K_{k+1}(r_{k+1},\tilde{d}_{k+1}) \tilde{A}(r_k,r_{k+1})\hat{x}_k\nonumber\\
&+ 2\,
\tilde{u}_k^T\tilde{B}^T(r_{k+1},d_k) K_{k+1}(r_{k+1},\tilde{d}_{k+1}) \sum\limits_{i=r_{k+1}}^{r_k}\bar{A}_{i}(r_{k+1},d_{k-i})\hat{x}_k\nonumber\\
&+
\hat{x}_k^T\tilde{A}^T(r_k,r_{k+1}) K_{k+1}(r_{k+1},\tilde{d}_{k+1})\tilde{A}(r_k,r_{k+1})\hat{x}_k\nonumber\\
&+ 2\,
\hat{x}_k^T\tilde{A}^T(r_k,r_{k+1}) K_{k+1}(r_{k+1},\tilde{d}_{k+1}) \sum\limits_{i=r_{k+1}}^{r_k}\bar{A}_{i}(r_{k+1},d_{k-i})\hat{x}_k\nonumber\\
&+
\hat{x}_k^T\sum\limits_{i=r_{k+1}}^{r_k}\bar{A}_{i}^T(r_{k+1},d_{k-i}) K_{k+1}(r_{k+1},\tilde{d}_{k+1}) \sum\limits_{j=r_{k+1}}^{r_k}\bar{A}_{j}(r_{k+1},d_{k-j})\hat{x}_k\nonumber
\end{align}
so $E_3$ from \eqref{eq:23} can be written as
\begin{align}
E_3 &= \mathbb{E}\left\{\left. 
v_{k+1} \right| \mathcal{I}_{k}\right\}\\
&= \hat{x}_k^T \hat{H}_{k}(r_k,\tilde{d}_k)\hat{x}_k + 2\,\tilde{u}_k^T\hat{M}_{k}(r_k,\tilde{d}_k)\hat{x}_k + \tilde{u}_k^T\hat{O}_{k}(r_k,\tilde{d}_k)\tilde{u}_k \label{eq:34}
\end{align}
with
\begin{align}
\hat{M}_{k}(r_k,\tilde{d}_k) &= \hat{M}_{1,k}(r_k,\tilde{d}_k) + \hat{M}_{2,k}(r_k,\tilde{d}_k)\\
\hat{H}_{k}(r_k,\tilde{d}_k) &= \hat{H}_{11,k}(r_k,\tilde{d}_k) + \hat{H}_{12,k}(r_k,\tilde{d}_k) + \hat{H}_{12,k}^T(r_k,\tilde{d}_k) + \hat{H}_{22,k}(r_k,\tilde{d}_k)\label{eq:36}
\end{align}
where, with $\tilde{\mathcal{D}}_{k+1} = \mathcal{D}_{k-\mathcal{R}_{k+1}}$,
\begin{align}
\hat{O}_{k}(r_k,\tilde{d}_k) &= \mathbb{E}\left\{\left. 
\tilde{B}^T(\mathcal{R}_{k+1},\mathcal{D}_k) K_{k+1}(\mathcal{R}_{k+1},\tilde{\mathcal{D}}_{k+1}) \tilde{B}(\mathcal{R}_{k+1},\mathcal{D}_k)
\,\right|\, \mathcal{I}_k\right\}\label{eq:35}\\
&=
\sum_{\rho=\underline{r}}^{\text{min}(\overline{r},1+r_k)}\mathbb{P}\left(\left. \mathcal{R}_{k+1}=\rho \,\right|\, \mathcal{I}_k\right)
\sum_{\delta_1=\underline{d}}^{\overline{d}}\mathbb{P}\left(\left. \mathcal{D}_{k-\rho}=\delta_1 \,\right|\, \mathcal{I}_k, \mathcal{R}_{k+1}=\rho\right)\\
&\qquad\qquad \sum_{\delta_2=\underline{d}}^{\overline{d}}\mathbb{P}\left(\left. \mathcal{D}_{k}=\delta_2 \,\right|\, \mathcal{I}_k, \mathcal{R}_{k+1}=\rho, \mathcal{D}_{k-\rho}=\delta_1\right)
\tilde{B}^T(\rho,\delta_2) K_{k+1}(\rho,\delta_1) \tilde{B}(\rho,\delta_2)\nonumber\\
&=
\sum_{\rho=\underline{r}}^{\text{min}(\overline{r},1+r_k)}\mathbb{P}\left(\left. \mathcal{R}_{k+1}=\rho \,\right|\, \mathcal{R}_k = r_k\right)
\sum_{\delta_1=\underline{d}}^{\overline{d}}\mathbb{P}\left(\left. \mathcal{D}_{k-\rho}=\delta_1 \,\right|\, \mathcal{D}_{k-1-r_k}=\tilde{d}_k\right)\\
&\qquad\qquad \sum_{\delta_2=\underline{d}}^{\overline{d}}\mathbb{P}\left(\left. \mathcal{D}_{k}=\delta_2 \,\right|\, \mathcal{D}_{k-\rho}=\delta_1\right)
\tilde{B}^T(\rho,\delta_2) K_{k+1}(\rho,\delta_1) \tilde{B}(\rho,\delta_2)\nonumber\\
&=
\sum_{\rho=\underline{r}}^{\text{min}(\overline{r},1+r_k)} \Psi(r_k,\rho)
\sum_{\delta_1=\underline{d}}^{\overline{d}}\Phi_{(1+r_k-\rho)}(\tilde{d}_k,\delta_1)
\sum_{\delta_2=\underline{d}}^{\overline{d}}\Phi_{(\rho)}(\delta_1,\delta_2)\\
&\qquad\qquad \tilde{B}^T(\rho,\delta_2) K_{k+1}(\rho,\delta_1) \tilde{B}(\rho,\delta_2)\nonumber
\end{align}
using \eqref{eq2_11} and \eqref{eq:P_7} since $r_k$ is the most recent available value in $\mathbf{r}$,  $d_{k-1-r_k} = \tilde{d}_{k}$ is the most recent available value in $\mathbf{d}$ and $k \geq k-\rho \geq k-1-r_k$ for $i\leq j$. Therefore also
\begin{align}
\hat{M}_{1,k}(r_k,\tilde{d}_k) &= \mathbb{E}\left\{\left. 
\tilde{B}^T(\mathcal{R}_{k+1},\mathcal{D}_k) K_{k+1}(\mathcal{R}_{k+1},\tilde{\mathcal{D}}_{k+1}) \tilde{A}(r_k,\mathcal{R}_{k+1})
\,\right|\, \mathcal{I}_k\right\}\\
&=
\sum_{\rho=\underline{r}}^{\text{min}(\overline{r},1+r_k)} \Psi(r_k,\rho)
\sum_{\delta_1=\underline{d}}^{\overline{d}}\Phi_{(1+r_k-\rho)}(\tilde{d}_k,\delta_1)
\sum_{\delta_2=\underline{d}}^{\overline{d}}\Phi_{(\rho)}(\delta_1,\delta_2)\\
&\qquad\qquad \tilde{B}^T(\rho,\delta_2) K_{k+1}(\rho,\delta_1) \tilde{A}(r_k,\rho)\nonumber
\end{align}
and
\begin{align}
\hat{M}_{2,k}(r_k,\tilde{d}_k) &= \mathbb{E}\left\{\left. 
\tilde{B}^T(\mathcal{R}_{k+1},\mathcal{D}_k) K_{k+1}(\mathcal{R}_{k+1},\tilde{\mathcal{D}}_{k+1}) \sum\limits_{i=\mathcal{R}_{k+1}}^{r_k}\bar{A}_{i}(\mathcal{R}_{k+1},\mathcal{D}_{k-i})
\,\right|\, \mathcal{I}_k\right\}\\
&=
\mathbb{E}\left\{\left. 
\sum\limits_{i=\mathcal{R}_{k+1}}^{r_k}\tilde{B}^T(\mathcal{R}_{k+1},\mathcal{D}_k) K_{k+1}(\mathcal{R}_{k+1},\tilde{\mathcal{D}}_{k+1}) \bar{A}_{i}(\mathcal{R}_{k+1},\mathcal{D}_{k-i})
\,\right|\, \mathcal{I}_k\right\}\\
&=
\sum_{\rho=\underline{r}}^{\text{min}(\overline{r},1+r_k)} \Psi(r_k,\rho)
\sum\limits_{i=\rho}^{r_k}
\sum_{\delta_1=\underline{d}}^{\overline{d}}\Phi_{(1+r_k-i)}(\tilde{d}_k,\delta_1)
\sum_{\delta_2=\underline{d}}^{\overline{d}}\Phi_{(i-\rho)}(\delta_1,\delta_2)\\
&\qquad\qquad \sum_{\delta_3=\underline{d}}^{\overline{d}}\Phi_{(\rho)}(\delta_2,\delta_3)
\tilde{B}^T(\rho,\delta_3) K_{k+1}(r_{k+1},\delta_2) \bar{A}_{i}(\rho,\delta_1)\nonumber
\end{align}
since also $k-1-r_k \leq k-i \leq k-r_{k+1}$. The same applies for
\begin{align}
\hat{H}_{11,k}(r_k,\tilde{d}_k) &= \mathbb{E}\left\{\left. \label{eq:37}
\tilde{A}^T(r_k,\mathcal{R}_{k+1}) K_{k+1}(\mathcal{R}_{k+1},\tilde{\mathcal{D}}_{k+1})\tilde{A}(r_k,\mathcal{R}_{k+1})
\,\right|\, \mathcal{I}_k\right\}\\
&=
\sum_{\rho=\underline{r}}^{\text{min}(\overline{r},1+r_k)} \Psi(r_k,\rho)
\sum_{\delta=\underline{d}}^{\overline{d}}\Phi_{(1+r_k-\rho)}(\tilde{d}_k,\delta)
\tilde{A}^T(r_k,\rho) K_{k+1}(\rho,\delta)\tilde{A}(r_k,\rho)
\end{align}
and
\begin{align}
\hat{H}_{12,k}(r_k,\tilde{d}_k) &= \mathbb{E}\left\{\left. \label{eq:38}
\tilde{A}^T(r_k,\mathcal{R}_{k+1}) K_{k+1}(\mathcal{R}_{k+1},\tilde{\mathcal{D}}_{k+1}) \sum\limits_{i=\mathcal{R}_{k+1}}^{r_k}\bar{A}_{i}(\mathcal{R}_{k+1},\mathcal{D}_{k-i})
\,\right|\, \mathcal{I}_k\right\}\\
&= \mathbb{E}\left\{\left. 
\sum\limits_{i=\mathcal{R}_{k+1}}^{r_k}\tilde{A}^T(r_k,\mathcal{R}_{k+1}) K_{k+1}(\mathcal{R}_{k+1},\tilde{\mathcal{D}}_{k+1}) \bar{A}_{i}(\mathcal{R}_{k+1},\mathcal{D}_{k-i})
\,\right|\, \mathcal{I}_k\right\}\\
&=
\sum_{\rho=\underline{r}}^{\text{min}(\overline{r},1+r_k)} \Psi(r_k,\rho)
\sum\limits_{i=\rho}^{r_k}
\sum_{\delta_1=\underline{d}}^{\overline{d}}\Phi_{(1+r_k-i)}(\tilde{d}_k,\delta_1)
\sum_{\delta_2=\underline{d}}^{\overline{d}}\Phi_{(i-\rho)}(\delta_1,\delta_2)\\
&\qquad\qquad \tilde{A}^T(r_k,\rho) K_{k+1}(\rho,\delta_2) \bar{A}_{i}(\rho,\delta_1).
\end{align}
\begin{align}
\hat{H}_{22,k}(r_k,\tilde{d}_k) &= \mathbb{E}\left\{\left. \label{eq:39}
\sum\limits_{i=\mathcal{R}_{k+1}}^{r_k}\bar{A}_{i}^T(\mathcal{R}_{k+1},\mathcal{D}_{k-i}) K_{k+1}(\mathcal{R}_{k+1},\tilde{\mathcal{D}}_{k+1}) \sum\limits_{j=\mathcal{R}_{k+1}}^{r_k}\bar{A}_{j}(\mathcal{R}_{k+1},\mathcal{D}_{k-j})
\,\right|\, \mathcal{I}_k\right\}\\
&= \mathbb{E}\left\{\left. 
\sum\limits_{i=\mathcal{R}_{k+1}}^{r_k}\sum\limits_{j=\mathcal{R}_{k+1}}^{r_k}\bar{A}_{i}^T(\mathcal{R}_{k+1},\mathcal{D}_{k-i}) K_{k+1}(\mathcal{R}_{k+1},\tilde{\mathcal{D}}_{k+1}) \bar{A}_{j}(\mathcal{R}_{k+1},\mathcal{D}_{k-j})
\,\right|\, \mathcal{I}_k\right\}\\
&=
\sum_{\rho=\underline{r}}^{\text{min}(\overline{r},1+r_k)} \Psi(r_k,\rho) \sum\limits_{i=\rho}^{r_k}\sum\limits_{j=\rho}^{r_k}
\sum_{\delta_1=\underline{d}}^{\overline{d}}\sum_{\delta_2=\underline{d}}^{\overline{d}}\sum_{\delta_3=\underline{d}}^{\overline{d}}\tilde{\mathcal{P}}(r_k,\tilde{d}_k,i,j,\rho,\delta_1,\delta_2,\delta_3)\\
&\qquad\qquad \bar{A}_{i}^T(\rho,\delta_1) K_{k+1}(\rho,\delta_3) \bar{A}_{j}(\rho,\delta_2)\nonumber
\end{align}
with
\begin{align}
&\tilde{\mathcal{P}}(r_k,\tilde{d}_k,i,j,\rho,\delta_1,\delta_2,\delta_3) \nonumber\\
&= 
\mathbb{P}\left(\left.\mathcal{D}_{k-i} = \delta_1\,\right.|\,\mathcal{I}_k\right)
\mathbb{P}\left(\left.\mathcal{D}_{k-j} = \delta_2\,\right.|\,\mathcal{I}_k, \mathcal{D}_{k-i} = \delta_1 \right)
\mathbb{P}\left(\left.\,\mathcal{D}_{k-\rho} = \delta_3\right.|\,\mathcal{I}_k, \mathcal{D}_{k-i} = \delta_1, \mathcal{D}_{k-j} = \delta_2 \right)\\
&= 
\mathbb{P}\left(\left.\mathcal{D}_{k-j} = \delta_2\,\right.|\,\mathcal{I}_k\right)
\mathbb{P}\left(\left.\mathcal{D}_{k-i} = \delta_1\,\right.|\,\mathcal{I}_k, \mathcal{D}_{k-j} = \delta_2 \right)
\mathbb{P}\left(\left.\,\mathcal{D}_{k-\rho} = \delta_3\right.|\,\mathcal{I}_k, \mathcal{D}_{k-j} = \delta_2, \mathcal{D}_{k-i} = \delta_1 \right)\\
&= 
\begin{cases}
\mathbb{P}\left(\left.\mathcal{D}_{k-i} = \delta_1\,\right.|\,\mathcal{D}_{k-1-r_k} = \tilde{d}_k\right)
\mathbb{P}\left(\left.\mathcal{D}_{k-j} = \delta_2\,\right.|\,\mathcal{D}_{k-i} = \delta_1 \right)
\mathbb{P}\left(\left.\,\mathcal{D}_{k-\rho} = \delta_3\right.|\, \mathcal{D}_{k-j} = \delta_2 \right)
& i\geq j\\
\mathbb{P}\left(\left.\mathcal{D}_{k-j} = \delta_2\,\right.|\,\mathcal{D}_{k-1-r_k} = \tilde{d}_k\right)
\mathbb{P}\left(\left.\mathcal{D}_{k-i} = \delta_1\,\right.|\,\mathcal{D}_{k-j} = \delta_2 \right)
\mathbb{P}\left(\left.\,\mathcal{D}_{k-\rho} = \delta_3\right.|\, \mathcal{D}_{k-i} = \delta_1 \right)
& i\leq j
\end{cases}\\
&= 
\begin{cases}
\Phi_{(1+i+r_k)}(\tilde{d}_k,\delta_1)\,\Phi_{(i-j)}(\delta_1,\delta_2)\,
\Phi_{(j-\rho)}(\delta_2,\delta_3)& i\geq j\\
\Phi_{(1+j+r_k)}(\tilde{d}_k,\delta_2)\,\Phi_{(j-i)}(\delta_2,\delta_1)
\,\Phi_{(i-\rho)}(\delta_1,\delta_3)& i\leq j
\end{cases}\\
&= 
\begin{cases}
\Phi_{(1+i+r_k)}(\tilde{d}_k,\delta_1)\,\Phi_{(i-j)}(\delta_1,\delta_2)\,
\Phi_{(j-\rho)}(\delta_2,\delta_3)& i>j\\
\Phi_{(1+j+r_k)}(\tilde{d}_k,\delta_2)\,\Phi_{(j-i)}(\delta_2,\delta_1)
\,\Phi_{(i-\rho)}(\delta_1,\delta_3)& i<j\\
\Phi_{(1+i+r_k)}(\tilde{d}_k,\delta_1)\,\Phi_{(i-\rho)}(\delta_1,\delta_3)& i=j,\delta_1=\delta_2\\
0& i=j,\delta_1\neq\delta_2
\end{cases}
\end{align}
using \eqref{eq2_11}, since \begin{itemize}
	\item $r_k$ is the most recent available value in $\mathbf{r}$,
	\item $d_{k-1-r_k} = \tilde{d}_{k}$ is the most recent available value in $\mathbf{d}$,
	\item $k-\rho \geq k-j\geq k-i \geq k-1-r_k$ for $i\geq j$,
	\item and $k-\rho \geq k-i\geq k-j \geq k-1-r_k$for $i\leq j$.
\end{itemize}
Since $\hat{H}_{11}(r_k,\tilde{d}_k) = \hat{H}^T_{11}(r_k,\tilde{d}_k)$ and $\hat{H}_{22}(r_k,\tilde{d}_k) = \hat{H}_{22}^T(r_k,\tilde{d}_k)$ also $\hat{H}(r_k,\tilde{d}_k) = \hat{H}^T(r_k,\tilde{d}_k)$.

With \eqref{eq:33}, \eqref{eq:29} and \eqref{eq:34}, \eqref{eq:32} can be written as
\begin{align}
v_{k} &= 
\underset{\tilde{u}_{k}}{\text{min }} \left(E_1 + E_2 + E_3\right)\\
&=
\underset{\tilde{u}_{k}}{\text{min}}\left[
\hat{x}_k^T \hat{H}_{k}(r_k,\tilde{d}_k)\hat{x}_k + 2\,\tilde{u}_k^T\hat{M}_{k}(r_k,\tilde{d}_k)\hat{x}_k + \tilde{u}_k^T\hat{O}_{k}(r_k,\tilde{d}_k)\tilde{u}_k\right.\\
&\qquad\qquad\left. +
\hat{x}_k^T \hat{Q}(r_k,\tilde{d}_k)\hat{x}_k + \tilde{u}_{k}^T \hat{R}(r_k,\tilde{d}_{k}) \tilde{u}_{k}
\right]\nonumber\\
&=
\underset{\tilde{u}_{k}}{\text{min}}\left[
\hat{x}_k^T \left(\hat{H}_{k}(r_k,\tilde{d}_k) + \hat{Q}(r_k,\tilde{d}_k) \right) \hat{x}_k
+ 2\,\tilde{u}_k^T\hat{M}_{k}(r_k,\tilde{d}_k)\hat{x}_k\right.\\
&\qquad\qquad\left. +
\tilde{u}_k^T \left( \hat{O}_{k}(r_k,\tilde{d}_k) + \hat{R}(r_k,\tilde{d}_{k}) \right) \tilde{u}_k
\right]\nonumber
\end{align}
which yields
\begin{align}
2\,\left( \hat{O}_{k}(r_k,\tilde{d}_k) + \hat{R}(r_k,\tilde{d}_{k}) \right) \tilde{u}_k
+ 2\,\hat{M}_{k}(r_k,\tilde{d}_k)\hat{x}_k = 0
\end{align}
by setting the first derivative with respect to $\tilde{u}_k$ of the term to be minimized to zero.\\

$\hat{O}_{k}(r_k,\tilde{d}_k)$ is symmetric and positive semi-definite since, as defined in \eqref{eq:35},
\begin{align}
\hat{O}_{k}(r_k,\tilde{d}_k) &= \mathbb{E}\left\{\left. 
\tilde{B}^T(\mathcal{R}_{k+1},\mathcal{D}_k) K_{k+1}(\mathcal{R}_{k+1},\tilde{\mathcal{D}}_{k+1}) \tilde{B}(\mathcal{R}_{k+1},\mathcal{D}_k)
\,\right|\, \mathcal{I}_k\right\}
\end{align}
where $K_{k+1}(\mathcal{R}_{k+1},\tilde{\mathcal{D}}_{k+1})$ is symmetric and positive semi-definite.\\

Since $\hat{O}_{k}(r_k,\tilde{d}_k) + \hat{R}(r_k,\tilde{d}_{k}) \succ 0$,
\begin{align}
\tilde{u}_k &= -\left( \hat{O}_{k}(r_k,\tilde{d}_k) + \hat{R}(r_k,\tilde{d}_{k}) \right)^{-1} \hat{M}_{k}(r_k,\tilde{d}_k)\hat{x}_k\\
&= -L_k((r_k,\tilde{d}_k))\hat{x}_k.
\end{align}\\

$\hat{H}_{k}(r_k,\tilde{d}_k)$ is symmetric and positive definite since inserting \eqref{eq:37}, \eqref{eq:38} and \eqref{eq:39} in \eqref{eq:36} can be written as
\begin{align}
\hat{H}_{k}(r_k,\tilde{d}_k) &= \hat{H}_{11,k}(r_k,\tilde{d}_k) + \hat{H}_{12,k}(r_k,\tilde{d}_k) + \hat{H}_{12,k}^T(r_k,\tilde{d}_k) + \hat{H}_{22,k}(r_k,\tilde{d}_k)\\
&=
\mathbb{E}\left\{\left.\left(
\tilde{A}(r_k,\mathcal{R}_{k+1}) + \sum\limits_{i=\mathcal{R}_{k+1}}^{r_k}\bar{A}_{i}(\mathcal{R}_{k+1},\mathcal{D}_{k-i})
\right)^T K_{k+1}(\mathcal{R}_{k+1},\tilde{\mathcal{D}}_{k+1}) \right.\right.\\
& \qquad\qquad\qquad\qquad\qquad \left.\left.\left(
\tilde{A}(r_k,\mathcal{R}_{k+1}) + \sum\limits_{i=\mathcal{R}_{k+1}}^{r_k}\bar{A}_{i}(\mathcal{R}_{k+1},\mathcal{D}_{k-i})
\right)\,\right|\, \mathcal{I}_k\right\}
\end{align}
where $K_{k+1}(\mathcal{R}_{k+1},\tilde{\mathcal{D}}_{k+1})$ is symmetric and positive semi-definite.\\

\begin{align}
v_{k} &= \hat{x}_k^T K_{k}(r_k,\tilde{d}_k)\hat{x}_k
\end{align}
with
\begin{align}
K_{k}(r_k,\tilde{d}_k) &=
\hat{H}_{k}(r_k,\tilde{d}_k) + \hat{Q}(r_k,\tilde{d}_k)
- 
\hat{M}_{k}^T(r_k,\tilde{d}_k) \left( \hat{O}_{k}(r_k,\tilde{d}_k) + \hat{R}(r_k,\tilde{d}_{k}) \right)^{-1} \hat{M}_{k}(r_k,\tilde{d}_k)
\end{align}
where $K_{k}(r_k,\tilde{d}_k) = K^T_{k}(r_k,\tilde{d}_k) \succeq 0$ since $\hat{H}_{k}(r_k,\tilde{d}_k) + \hat{Q}(r_k,\tilde{d}_k) = \hat{H}^T_{k}(r_k,\tilde{d}_k) + \hat{Q}^T(r_k,\tilde{d}_k) \succeq 0$ and $\hat{O}_{k}(r_k,\tilde{d}_k) + \hat{R}(r_k,\tilde{d}_{k}) = \hat{O}^T_{k}(r_k,\tilde{d}_k) + \hat{R}^T(r_k,\tilde{d}_{k}) \succ 0$.

%%%%%%%%%%%%%%%%%%%%%%%%%%%%%%%%%%%%%%%%%%%%%%%%%%%%%%%%%%%%%%%%%%%%%%%%%%%%%%%%%%
\newpage
%\printbibliography

\end{document}